\documentclass[twoside,11pt, english]{article}
\usepackage[top=2cm, bottom=2cm, outer=2.5cm, inner=2.5cm]{geometry}
\usepackage[utf8]{inputenc}
\usepackage{amsmath}
\usepackage{physics} 
\usepackage{amsthm}
\usepackage{amsfonts}
\usepackage{amssymb}
\usepackage{graphicx}
\usepackage[dvipsnames]{xcolor}
\usepackage{hyperref} 
\usepackage{mathrsfs}
\usepackage{authblk}
\usepackage{tikz}
\usepackage[nameinlink]{cleveref}
\usetikzlibrary{calc,decorations.markings}
\usetikzlibrary{decorations.pathmorphing}

\definecolor{ClaColor}{RGB}{0,0,255}

\title{Ground State for the Klein-Gordon field in anti-de Sitter spacetime with dynamical Wentzell boundary conditions}
\author{Claudio Dappiaggi$^{1,2}$\thanks{\href{mailto:claudio.dappiaggi@unipv.it}{claudio.dappiaggi@unipv.it}}, Benito A. Ju\'arez-Aubry$^{3}$\thanks{\href{mailto:benito.juarez@iimas.unam.mx}{benito.juarez@correo.nucleares.unam.mx}} and Alessio Marta$^{4,5}$\thanks{\href{mailto:alessio.marta@unimi.it}{alessio.marta@$•$unimi.it}}}

\newcommand{\beq}{\begin{equation}}
	\newcommand{\ene}{\end{equation}}

\newtheorem{thm}{Theorem}

\newtheorem{rem}[thm]{Remark}

\affil{$^{1}$Dipartimento  di  Fisica,  Universit\`a degli Studi di Pavia, Via  Bassi,  6,  27100  Pavia,  Italy}  
\affil{$^{2}$Istituto Nazionale di Fisica Nucleare --  Sezione di Pavia, Via Bassi, 6, 27100 Pavia, Italy}
\affil{$^{3}$Departamento de Gravitaci\'on y Teor\'ia de Campos, Instituto de Ciencias Nucleares, Universidad Nacional Aut\'onoma de M\'exico, A. Postal 70-543, Mexico City 045010, Mexico}
\affil{$^{4}$Dipartimento  di  Matematica,  Universit\`a  degli  Studi  di  Milano,
	Via  Saldini  50,  20133  Milano,  Italy}
\begin{document}
	
	\date{}
	
	\maketitle
	
	\begin{abstract}
We consider a real Klein-Gordon field in the Poincar\'e patch of $(d+1)$-dimensional anti-de Sitter spacetime, PAdS$_{d+1}$, and impose dynamical boundary condition on the asymptotic boundary of PAdS$_{d+1}$ that depend explicitly on the second time derivative of the field at the boundary. These boundary conditions are of generalized Wentzell type. We construct the Wightman two-point function for the ground state of the Klein-Gordon theory whenever the parameters of the theory (the field mass, curvature coupling and boundary condition parameters) render such ground state admissible. In the cases in which the mass of the Klein-Gordon field and the curvature coupling term yield an effectively massless theory, we can define a boundary field whose dynamics are ruled by the dynamical boundary condition and construct, in addition to the Wightman function for the Klein-Gordon field, boundary-to-boundary, boundary-to-bulk and bulk-to-boundary propagators.
	\end{abstract}

\section{Introduction}\label{Sec: Intro}

Interest in studying classical and quantum field theory in (asymptotically) anti-de Sitter spacetimes (AdS) has steadily increased in the last decades for several reasons. On the one hand, the remarkable AdS/CFT conjecture \cite{Maldacena}, together with statements on more general holographic dualities, has added interest to analyzing physical situations in (asymptotically) AdS and other spacetimes with (asymptotic) timelike boundaries. On the other hand, while our understanding of classical and quantum field theory in globally hyperbolic spacetimes has advanced remarkably in the past decades, the situation is less developed for spacetimes with (asymptotic) timelike boundaries.

One of the reasons for this is that global well-posedness in spacetimes with timelike boundaries or asymptotic timelike boundaries \cite{AFS18}, such as AdS, requires prescribing boundary conditions in addition to Cauchy data. Moreover, infinitely many boundary conditions will give rise to reasonable (classical and quantum) field theories. For example, quantum field theory in AdS spacetimes with Robin boundary conditions has been studied in \cite{DF16, Dappiaggi:2017wvj, DFM18} and the analogous problem for Lifshitz spacetimes has been addressed in \cite{DeSouzaCampos:2021dma}.

The purpose of this paper is to study quantum field theory in AdS spacetimes with {\it dynamical Wentzell boundary conditions} (WBC), thus extending the analysis that has been carried out in \cite{DFJ18} where the classical mode solutions were obtained for a real Klein-Gordon equation in this case. These dynamical boundary conditions are such that a condition is imposed on the second time derivative of the field at the boundary. In other words, the ensuing system codifies on the one hand the bulk dynamics, while, on the other hand, it reads boundary observables as suitable traces of bulk fields, subject to an evolution dictated by the boundary conditions.

There are a number of  motivations for studying Wentzell boundary conditions in AdS, especially from the quantum field theoretic viewpoint. First of all, it is clear that a system where bulk and boundary observables can be naturally defined in AdS is attractive as a simple model reminiscent of AdS/CFT, but in which all mathematical subtleties are under control. In addition these boundary conditions are also closely related to the ones appearing in so-called holographic renormalization \cite{Skenderis}. As mentioned in the introductions of \cite{DFJ18, G.:2015yxa, Barbero:2017kvi}, good motivations to introduce these boundary conditions stem from the study of condensed matter systems and isolated horizons in gravitation. Perhaps one of the most important instances in which this boundary conditions are relevant are in the experimental verification of the dynamical Casimir effect \cite{Wilson:2011}, see \cite{Juarez-Aubry:2021qfv} for an overview.

Additionally, as explained in \cite{DFJ18} WBC are distinguished also mathematically in that they generalize Robin boundary conditions and are compatible with the spacetime isometries, thus providing good candidate systems for constructing ground states in quantum field theory. Additionally, they provide a generalization to the classical work of Ishibashi and Wald \cite{Ishibashi:2004wx} in which the boundary conditions in AdS are seen as self-adjoint extensions of certain elliptic operators on $L^2(\Sigma)$, the space of square integrable functions on a Cauchy surface, identified with respect to the standard AdS time coordinate. Indeed, it turns out that the natural Hilbert space for systems with dynamical boundary conditions has to be extended so to be able to account also for dynamical boundary degrees of freedom, see \cite{Zahn:2015due}, but also \cite{Weder1, Weder2}. In these papers the analysis is inspired by the so-called boundary eigenvalue problems, see e.g. \cite{Mennicken}. Moreover, it has been shown in \cite{Weder2} that there are systems with WBC that are F-local quantum field theories in the sense of Kay \cite{Kay:1992es}.

With these motivations, the concrete goal of the present paper is to construct the ground state Wightman two-point function for a class of real Klein-Gordon theories in the Poincar\'e patch of anti-de Sitter spacetime. From the point of view of algebraic quantum field theory, this completely characterizes the Klein-Gordon theory in the spacetime of interest via the GNS theorem, whereby a concrete representation of operators on a Hilbert space can be constructed. A major point of interest is that, as we shall see, in the case in which the scalar field combined with the AdS curvature coupling yield an effectively massless theory, it is also possible to assign a two-point function for the boundary theory confined to the AdS timelike boundary, as well as bulk-to-boundary and boundary-to-bulk propagators in addition to the bulk Wightman function. An important aspect of Wightman two-point functions is that their antisymmetric part coincides up to a multiplicative constant with the advanced-minus-retarded fundamental solution of the Klein-Gordon operator. On a globally hyperbolic spacetime without boundary, this feature is known to carry two bits of information. On the one hand it codifies the canonical commutation relations on any constant time, Cauchy hypersurface. On the other hand the axiom of causality, proper of relativistic quantum field theories, is translated in the property that the image of the advanced and of the retarded fundamental solutions is supported respectively in the past and in the future light cone of the test-function. This second statement is no longer automatically true in presence of a timelike boundary such as when we consider asymptotically AdS spacetimes as background. In this case the r\^{o}le of boundary conditions entails that all support properties must be verified form scratch and, to the best of our knowledge, a general proof does not exist. Yet, using energy estimates, in this paper we are able to prove the support properties for the fundamental solutions of a massless Klein-Gordon field on the Poincar\'e patch of an AdS spacetime of arbitrary dimension, endowed with Wentzell boundary conditions.

The organization of the paper is as follows. In Sec. \ref{Sec: Data} we will introduce the basic geometric ideas of the Poincar\'e patch of AdS (PAdS) and the Klein-Gordon equation in this spacetime. This will also serve the purpose of fixing the notation of the subsequent sections. The Wightman function for the bulk Klein-Gordon field in PAdS will be constructed in Sec. \ref{Sec:GroundState}. The key point will be to use the symmetry of spacetime to rewrite the underlying equation of motion as a one-dimensional Sturm-Liouville problem. The question of obtaining the Wightman function then reduces to obtaining the resolvent operator of the relevant Sturm-Liouville operator. We include a discussion on the support properties of the causal propagator of the theory in Sec. \ref{sec:Support}. In particular, for an effectively massless theory we can show that the support properties are as expected, but a result for more general theories remains elusive and is indeed an important open question. Our final remarks appear in Sec. \ref{sec:Rem}. 

\section{Preliminary Data}\label{Sec: Data}
In this short section we introduce the basic geometric and analytic data that will be useful in the following sections.

\paragraph{Geometry of AdS spacetimes --}

We consider the $(d+1)$-dimensional anti-de Sitter spacetime AdS$_{d+1}$, that is the maximally symmetric solution of Einstein equations with negative cosmological constant $\Lambda$. As a manifold AdS$_{d+1}$ is diffeomorphic to $\mathbb{S}^1 \times \mathbb{R}^d$ and it can be realized as an embedded submanifold in the pseudo-Riemannian manifold $(\mathbb{R}^{d+2},g_{d+2})$, where, considering the standard Cartesian coordinates $\{X_j\}_{j=0,\dots,d+1}$, the line element associated to $g_{d+2}$ reads $ds^2_{g_{d+2}} = -dX_0^2-dX_1^2 + \sum\limits_{j=2}^{d+1} dX_j^2$. In this context AdS$_{d+1}$ can be obtained imposing the constraint $-X_0^2-X_1^2 + \sum\limits_{j=2}^{d+1} X_j^2 = -\ell^2 $, where $\ell^2\doteq -\frac{d(d-1)}{\Lambda}$ is known as the radius of curvature of AdS$_{d+1}$. 

In this work we shall consider the {\em Poincar\'e fundamental domain/patch,} of AdS$_{d+1}$, denoted as PAdS$_{d+1}$, which can be represented in terms of the coordinates $(z,t,x_1,\cdots,x_{d-1})$, with $t,x_i \in \mathbb{R}$, $i=1,\dots,d-1$ while $z \in \mathbb{R}^+$, by means of the transformation
\begin{equation}\label{Eq: Poincare coordinates}
	\begin{cases}
		X_0 = \dfrac{\ell}{z}t\\
		X_1 = \dfrac{z}{2}\left(1+\dfrac{1}{z^2}\left(-t^2+\delta^{ij}x_ix_j+\ell^2\right)\right) \\
		X_i = \dfrac{\ell}{z}x_{i-1}, \ \ \ i = 2,\cdots,d\\
		X_{d+1} = \dfrac{z}{2}\left(1+\dfrac{1}{z^2}\left(-t^2+\delta^{ij}x_ix_j-\ell^2\right)\right) \\
	\end{cases}.
\end{equation}
One can realize by direct inspection that $X_1+X_{d+1}=\frac{\ell}{z}>0$ and hence the Poincar\'e patch covers only a portion of the whole AdS$_{d+1}$-spacetime. As a consequence of Equation \eqref{Eq: Poincare coordinates}, as a manifold PAdS$_{d+1}$ is isometric to $(\mathbb{R}^+ \times \mathbb{R}^d,g_{\textrm{PAdS}_{d+1}})$, while the metric $g_{\textrm{PAdS}_{d+1}}$ has the following line element
\begin{equation}\label{Eq: PAdS metric}
	ds^2_{PAdS_{d+1}} = \dfrac{\ell^2}{z^2} \left( -dt^2 + dz^2 + \sum\limits_{i=1}^{d-1} dx_i^2 \right).
\end{equation}
As one can observe directly from Equation \eqref{Eq: PAdS metric} $(\textrm{PAdS}_{d+1},g_{\textrm{PAdS}_{d+1}})$ is conformal to $\mathring{\mathbb{H}}^{d+1}$, the interior of the upper half space $\mathbb{H}^{d+1}$ realized as the subset of coordinates $z\geq 0$ of the $(d+1)$-dimensional Minkowski spacetime $(\mathbb{R}^{d+1},\eta)$. Hence PAdS$_{d+1}$ is a globally hyperbolic spacetime with timelike boundary, see \cite{AFS18} and the locus $z=0$, {\it i.e.} $\partial\mathbb{H}^{d+1}$, represents its conformal boundary. For convenience and without loss of generality, henceforth we set $\ell=1$.

\paragraph{The Klein-Gordon equation}
On top of ${\rm PAdS}_{d+1}$ we consider a real scalar field $\phi: \textrm{PAdS}_{d+1} \to \mathbb{R}$ satisfying the Klein-Gordon equation
\begin{align}
	P \phi := (\Box_g - m^2_0 - \xi R) \phi = 0.
	\label{KG-PAdS}
\end{align}
where $\Box_g$ is the d'Alembert wave operator built out of the metric as per Equation \eqref{Eq: PAdS metric}, $m_0$ is the mass parameter while $\xi \in \mathbb{R}$ is the coupling to the scalar curvature $R=-d(d+1)$. For later convenience we introduce an effective mass parameter 
\begin{equation}\label{Eq: effective mass}
	m^2\doteq m^2_0+\left(\xi-\frac{d-1}{4d}\right) R,
\end{equation}
which we require to abide to the Breitenlohner-Freedman bound $m^2\geq -\frac{1}{4}$ \cite{BF82}. 

Since  PAdS$_{d+1}$ is a globally hyperbolic spacetime with a timelike boundary in the sense formalized in \cite{AFS18}, once smooth and compactly supported initial data are assigned on a Cauchy surface, a unique solution of Equation \eqref{KG-PAdS} exists in their domain of dependence provided that this does not intersect the conformal boundary. On the contrary, if one is interested in global solutions, it is necessary to assign in addition boundary conditions at $z=0$. This problem has been thoroughly investigated in the past few years starting from the early work of Ishibashi and Wald \cite{Ishibashi:2004wx}, see also  
\cite{DF16,Dappiaggi:2017wvj,DFM18,DDF18,Pitelli:2019svx}.

In these papers, a great deal of attention has been reserved to studying the r\^{o}le of boundary conditions of Robin type, although they do not exhaust the set of all possible choices. As mentioned in the introduction, a particularly interesting, alternative option is represented by the so-called {\em generalized Wentzell}, dynamical boundary conditions, see e.g. \cite{DFJ18, G.:2015yxa, Barbero:2017kvi, Weder1, Weder2, Zahn:2015due}. In the following we introduce them, summarizing succinctly the analysis of \cite{DFJ18}. 

We consider Equation \eqref{KG-PAdS} with $m^2\geq 0$.
If we introduce the conformally rescaled field $\Phi=\Omega^{\frac{1-d}{2}}\phi:\mathring{\mathbb{H}}^{d+1}\to\mathbb{R}$ with $\Omega=z$, the Klein-Gordon equation with Wentzell boundary conditions reads 
\begin{align}
	\left\{
	\begin{array}{l} 
		P_\eta \Phi := \left(\Box_\eta - \frac{m^2}{z^2} \right) \Phi = 0,  \\
		\left(\Box_\eta^{(d)} - m_b^2 \right)  F = -\frac{\rho}{c}, \\
		\gamma_0(\Phi):=\lim\limits_{z \rightarrow 0}W_z[\Phi,\Phi_1] = F , \ \ \ \ \gamma_1(\Phi)=\lim\limits_{z \rightarrow 0}W_z[\Phi,\Phi_2] = \rho
	\end{array} \right., 
	\label{eq:KG_boundary_problem_singular}
\end{align}
where $\Box_\eta$ ({\em resp.} $\Box_\eta^{(d)}$) is the D'Alembert wave operator built out of the Minkowski metric on $\mathbb{H}^{d+1}$ ({\em resp.} $\mathbb{R}^d$), $m_b \geq 0$ can be interpreted as the mass of the boundary field $F$. In addition, given two arbitrary, differentiable functions $u$ and $v$ on $\mathring{\mathbb{H}}^{d+1}$, $ W_z [u,v]= u \partial_z v - v \partial_z u $ is the Wronskian between $u$ and $v$. 

Still focusing on Equation \eqref{eq:KG_boundary_problem_singular}, one needs a rationale to select the functions $\Phi_1$ and $\Phi_2$. To this end, observe that we can take the Fourier transform of all bulk and boundary fields along the directions tangent to $\partial\mathbb{H}^{d+1}$, rewriting Equation \eqref{eq:KG_boundary_problem_singular} as the following Sturm-Liouville problem:

\begin{subequations}
	\begin{equation}\label{eq:Sturm_Liouville_Singular}
		\left( -\partial_z^2 + \frac{m^2}{z^2} \right) \widehat{\Phi}(z,\underline{k}) = q^2 \widehat{\Phi}(z,\underline{k}), \hspace{8pt} z \in \mathbb{R}^+
	\end{equation}
	\begin{equation}\label{eq:Sturm_Liouville_Singular_boundary_conditions}
		m_b^2 \widehat{F}(\underline{k}) -\frac{1}{c} \widehat{\rho}(\underline{k}) = q^2  \widehat{F}(\underline{k}),
	\end{equation}
	\begin{equation}\label{eq:Sturm_Liouville_Singular_Wentzell_boundary_conditions}
		\lim\limits_{z \rightarrow 0}W_z[\widehat{\Phi}(z,\underline{k}),\widehat{\Phi}_1(z,\underline{k})] = \widehat{F}(\underline{k}) , \ \ \ \ \lim\limits_{z \rightarrow 0}W_z[\widehat{\Phi}(z,\underline{k}),\widehat{\Phi}_2(z,\underline{k})] = \widehat{\rho}(\underline{k}),
	\end{equation} 
\end{subequations}
where $\underline{k}=(\omega,k_1,\dots,k_{d-1})$, while $q^2=\omega^2-\sum\limits_{i=1}^{d-1}k^2_i$.
Using the standard nomenclature of Sturm-Liouville problems, see \cite{Zettl:2005} or \cite{DF16} for a short survey, we call $\widehat{\Phi}_1(z,\underline{k})$ the principal solution, namely the unique solution of Equation \eqref{eq:Sturm_Liouville_Singular} - up to scalar multiples - such that $\lim_{z \rightarrow 0} \widehat{\Phi}_1(\underline{k},z)/\widehat{\Psi}(\underline{k},z)=0$ for every solution $\widehat{\Psi}$ which is not a multiple of $\widehat{\Phi}_1$. On the contrary $\widehat{\Phi}_2(z,\underline{k})$ is any other, arbitrary but fixed, solution of Equation \eqref{eq:Sturm_Liouville_Singular} which is linearly independent from $\widehat{\Phi}_1(z,\underline{k})$.

In \cite{DFJ18}, the mode solutions associated to Equation \eqref{eq:Sturm_Liouville_Singular}, subject to the boundary condition as per Equation \eqref{eq:Sturm_Liouville_Singular_boundary_conditions}, have been studied in detail. We consider only the regime $c<-\nu^{\nu}\left( m_b^2/(1-\nu) \right)^{\nu-1}$, where $\nu := \sqrt{1+4 m^2}/2$, which guarantees that the operator $\widehat{A}_m$ possesses for all $m>0$ only a continuous spectrum, $q^2>0$ and no bound state occurs, \cite{DFJ18}. More precisely it turns out that Equation \eqref{eq:Sturm_Liouville_Singular} can be solved as
\begin{equation} \label{eq:solution_massive}
	\widehat{\Phi}(\underline{k},z) = \widehat{\rho}(\underline{k})\widehat{\Phi}_1(\underline{k},z)-\widehat{F}(\underline{k})\widehat{\Phi}_2(\underline{k},z),
\end{equation}
where
\begin{subequations}
	\begin{equation}\label{eq:primary_solution}
		\widehat{\Phi}_1(\underline{k},z) = \sqrt{\frac{\pi}{2}} q^{-\nu} \sqrt{z} J_\nu (qz),
	\end{equation}
while
	\begin{equation}\label{eq:secondary_solution}
		\widehat{\Phi}_2(\underline{k},z) =
		\begin{cases}
			-\sqrt{\frac{\pi z}{2}} q^{\nu} J_{-\nu} (qz) & \nu > 0,\\
			-\sqrt{\frac{\pi z}{2}} \left[ Y_0(qz) - \frac{2}{\pi} log(q) \right] & \nu = 0,
		\end{cases}
	\end{equation}
\end{subequations}
where $J_\nu$ and $Y_\nu$ represent the Bessel function of order $\nu$ respectively of the first and of the second kind. For later convenience we stress that, given any $z_0\in(0,\infty)$, $\widehat{\Phi}_1(k,z)\in L^2((0,z_0);dz)$, while $\widehat{\Phi}_2(k,z)\in L^2((0,z_0);dz)$ if and only if $\nu<1$. 

\begin{rem}
	We reckon that, for a reader who is not familiar with the language of Sturm-Liouville equations and of the associated boundary value problems, it is instructive to have a quick and closer look at the scenario where the bulk mass of the field vanishes, {\it i.e.}, $m=0$ or equivalently $\nu=\frac{1}{2}$. In this instance Equation \eqref{eq:KG_boundary_problem_singular} becomes an eigenvalue problem for the kinetic operator, namely
	$$-\partial_z^2 \widehat{\Phi}(z) = q^2 \widehat{\Phi}(z). \hspace{8pt} z \in \mathbb{R}^+ $$
While Equation \eqref{eq:Sturm_Liouville_Singular_boundary_conditions} is left unchanged, the principal and the secondary solutions as per Equation \eqref{eq:primary_solution} and \eqref{eq:secondary_solution} become
$$\widehat{\Phi}_1(\underline{k},z)=\frac{\sin(qz)}{q}\qquad\widehat{\Phi}_2(\underline{k},z)=\cos(qz).$$
This entails that one can give a more transparent interpretation to Equation \eqref{eq:Sturm_Liouville_Singular_Wentzell_boundary_conditions}, namely a direct calculation shows that 
$$\widehat{\Phi}(0,\underline{k})=\widehat{F}(\underline{k}),\quad(\partial_z\widehat{\Phi})(0,\underline{k})=\widehat{\rho}(\underline{k}),$$
which in turn entails that 
$$\gamma_0(\Phi)=\left.\Phi\right|_{z=0},\quad\gamma_1(\Phi)=\left.\partial_z\Phi\right|_{z=0}.$$
To summarize, if $m=0$, we are dealing with a so-called {\em regular Sturm-Liouville problem} and the boundary fields $F$ and $\rho$ can be read as the restriction to $\partial\mathbb{H}^{d+1}$ respectively of the bulk field and of its derivative along the direction normal to the boundary.
\end{rem}

\section{Ground State with Wentzell boundary conditions}
\label{Sec:GroundState}

In this section we discuss the existence of the Wightmann bulk-to-bulk two-point correlation function for a real massive scalar field on PAdS$_{d+1}$ with Wentzell boundary conditions. In other words we seek a bi-distribution $\lambda_2\in\mathcal{D}^\prime(\textrm{PAdS}_{d+1}\times\textrm{PAdS}_{d+1})$ such that the following three conditions are met:
\begin{description}
	\item[Dynamics] $(P\otimes\mathbb{I})\lambda_2=(\mathbb{I}\otimes P)\lambda_2=0$ where $P$ is the Klein-Gordon operator as per Equation \eqref{KG-PAdS}.
	\item[Positivity] $\lambda_2(f,f)\geq 0$ for any $f\in C^\infty_0(\textrm{PAdS}_{d+1})$,
	\item[CCR] for any $f,f^\prime\in C^\infty_0(\textrm{PAdS}_{d+1})$, $\lambda_2(f,f^\prime)-\lambda_2(f^\prime,f)=i G(f,f^\prime)$, 
\end{description}
where $G=G^+-G^-$, dubbed {\em causal propagator}, is the difference between the retarded $(+)$ and the advanced $(-)$ fundamental solutions of $P$ on PAdS$_{d+1}$, once a choice of boundary conditions has been made. In addition, among the plethora of existing two-point functions, we require that $\lambda_2$ is of Hadamard form. This is a constraint on the singular structure of $\lambda_2$ which codifies, on the one hand, that the ultraviolet behaviour of the underlying quantum state mimics that of the Poincar\'e-invariant Minkowski vacuum while, on the other hand, it entails the finiteness of the quantum fluctuations of all observables. First introduced and developed under the assumption that the underlying spacetime is globally hyperbolic, see \cite{Khavkine:2014mta} for a review, the Hadamard condition has been recently adapted to asymptotically AdS spacetimes, \cite{Dappiaggi:2017wvj,GaWr18,Wro17}. While the existence of Hadamard states is a well-established result on globally hyperbolic spacetimes thanks to a deformation argument \cite{FNW81}, the same conclusion cannot be drawn if the background possesses a timelike boundary unless the underlying boundary condition is time independent \cite{Dappiaggi:2021wtr}. On the contrary, when boundary conditions of Wentzell type are imposed, the question of existence of Hadamard states is still open. Here we shall address it under the assumption that the underlying background is the Poincar\'e patch of a $(d+1)$-dimensional AdS spacetime. In this endeavor, we must start by constructing a ground state. We shall divide the analysis in several steps.

\paragraph{Step 1: Reduction to $\mathbf{\mathring{\mathbb{\bf H}}^{d+1}}$ --} Following the rationale of Section \ref{Sec: Data}, it is convenient to consider a conformally related problem, namely we look for $\lambda_2^{\mathbb{H}}\in\mathcal{D}^\prime(\mathring{\mathbb{H}}^{d+1}\times\mathring{\mathbb{H}}^{d+1})$ such that, working at the level of integral kernel 
$$\lambda_2(x,x^\prime)=(zz^\prime)^{\frac{d-1}{2}}\lambda_2^{\mathbb{H}}(x,x^\prime).$$
In addition the three defining conditions for a two-point correlation function translate to
\begin{subequations}
	\begin{equation}
	(P_\eta\otimes\mathbb{I})\lambda^{\mathbb{H}}_2=(\mathbb{I}\otimes P_\eta)\lambda^{\mathbb{H}}_2=0,
	\end{equation}
where $P_\eta$ is defined in Equation \eqref{eq:KG_boundary_problem_singular},
\begin{equation}
	\lambda_2^{\mathbb{H}}(f,f)\geq 0,\quad \forall f\in C^\infty_0(\mathring{\mathbb{H}}^{d+1}),
\end{equation}
\begin{equation}
	\lambda_2^{\mathbb{H}}(f,f^\prime)-\lambda_2^{\mathbb{H}}(f^\prime,f)=i G^{\mathbb{H}}(f,f^\prime),\quad \forall f,f^\prime\in C^\infty_0(\mathring{\mathbb{H}}^{d+1}),
\end{equation}
where $G^{\mathbb{H}}$ is the causal propagator associated to the operator $P_\eta$ with Wentzell boundary conditions.
\end{subequations}

\paragraph{Step 2: Construction of the causal propagator $G^{\mathbb{H}}$ --} The next step consists on focusing on $G^{\mathbb{H}}$ constructing it explicitly starting from the mode solutions discussed in Section \ref{Sec: Data}. Observe that, a priori, the existence of advanced and retarded fundamental solutions for a normally hyperbolic operator is guaranteed if the underlying spacetime is globally hyperbolic. On the contrary, in presence of a timelike boundary such as in the case in hand, a separate analysis is necessary. In \cite{DDF18} it has been shown by means of purely functional analytic tools that existence is guaranteed when considering Wentzell boundary conditions, though only in the regular case.

Hence, in the following we shall exploit the mode solutions introduced in Section \ref{Sec: Data} to construct explicitly the causal propagator $G^{\mathbb{H}}$ associated to Equation \eqref{eq:KG_boundary_problem_singular}. Following the same rationale adopted in \cite{DF16,DFM18} and working at the level of integral kernel, $G^{\mathbb{H}}(x,x^\prime)$ is a solution of the initial value problem
\begin{equation}\label{eq:commutator_function_sturm_liouville}
	\begin{cases}
		(P_\eta \otimes \mathbb{I})G^{\mathbb{H}} = (\mathbb{I} \otimes P_\eta)G^{\mathbb{H}}= 0\\
		G^{\mathbb{H}}(x,x^\prime)|_{t=t^\prime} = 0\\
		\partial_t G^{\mathbb{H}}(x,x^\prime)|_{t=t^\prime} = - \partial_{t^\prime} G^{\mathbb{H}}(x,x^\prime) |_{t=t^\prime} = \delta(z-z^\prime)\prod\limits_{i=1}^{d-1}\delta(x_i-x^\prime_i)
	\end{cases},
\end{equation}
supplemented with the Wentzell boundary conditions. In view of translation invariance along all directions barring the one orthogonal to the boundary $\partial\mathbb{H}^{d+1}$, we can use the Fourier-Bessel transform to write
\begin{equation}\label{Eq: form_of_the_causal_propagator}
	G^{\mathbb{H}}(x,x^\prime)= \lim\limits_{\varepsilon \rightarrow 0^+} \int\limits_{\mathbb{R}} d\omega\sqrt{\dfrac{2}{\pi}} \frac{\sin(\omega(t-t^\prime-i\varepsilon))}{\omega}\int_0^\infty dk k \left(\dfrac{k}{r} \right)^{\frac{d-3}{2}} J_{\frac{d-3}{2}}(kr)\widehat{G}^{\mathbb{H}}_{k}(z,z^\prime),
\end{equation}
where we set $r^2 \doteq \sum\limits_{i=1}^{d-1}(x^i-x^{\prime i})^2$ while $k^2=\sum\limits_{i=1}^{d-1}k^2_i$, $k_i$ being the Fourier parameter associated to $x_i$. Equation \eqref{eq:commutator_function_sturm_liouville} entails that the only unknown $\widehat{G}^{\mathbb{H}}_{k}(z,z^\prime)$ is a solution of the eigenvalue problem
\begin{equation}\label{Eq: ODE_G}
	(L \otimes \mathbb{I})\widehat{G}^{\mathbb{H}}_{k}(z,z^\prime) = (\mathbb{I} \otimes L)  \widehat{G}^{\mathbb{H}}_{k}(z,z^\prime) = \lambda \widehat{G}^{\mathbb{H}}_{k}(z,z^\prime),\quad L=-\partial^2_z+\frac{m^2}{z^2},
\end{equation}
where the r\^{o}le of the spectral parameter is played by $\lambda = q^2 = \omega^2 - k^2$. The initial conditions yield the constraint
\begin{equation}\label{eq:delta-rep}
	\frac{(2 \pi)^{\frac{d}{2}} \Gamma \left(\frac{d-1}{2} \right)}{\sqrt{\pi}
		\Gamma \left( \frac{d}{2} \right)} \int_0^\infty dq\, q \widehat{G}^{\mathbb{H}}_{k}(z,z^\prime) = \delta(z-z^\prime),
\end{equation}
where we have implicitly assumed that $\widehat{G}^{\mathbb{H}}_{k}(z,z^\prime)$ does depend on the momenta only via $q$ and where we used the identity 
\begin{equation}
	\int_0^\infty dk k \left( \dfrac{k}{r} \right)^{\frac{d-3}{2}} J_{\frac{d-3}{2}}(kr) = \dfrac{(2\pi)^{\frac{d}{2}}\Gamma\left(\frac{d-1}{2}\right)}{\sqrt{2}\Gamma\left(\frac{d}{2} \right)}\prod_{i=1}^{d-1}\delta(x_i-x^\prime_i).
\end{equation}

Equation \eqref{eq:delta-rep} in combination with Equation \eqref{Eq: ODE_G} entails that we can construct $\widehat{G}^{\mathbb{H}}(z,z^\prime)$ starting from a resolution of the identity operator in terms of eigenfunctions of $L$. These are nothing but the mode solutions introduced in Section \ref{Sec: Data}. This is a procedure which has been already followed in \cite{DF16,DFM18} when constructing the ground state for a massive real scalar field on an AdS spacetime with Robin boundary conditions. The first step consists of constructing the so-called {\em radial Green function} $\mathcal{R}$ which obeys the following defining equation
$$(L\otimes\mathbb{I})\mathcal{R}=(\mathbb{I}\otimes L)\mathcal{R}=\delta(z,z^\prime).$$ 
Here we work once more at the level of integral kernels and we are assuming implicitly Wentzell boundary conditions as per Equation \eqref{eq:Sturm_Liouville_Singular_Wentzell_boundary_conditions}.

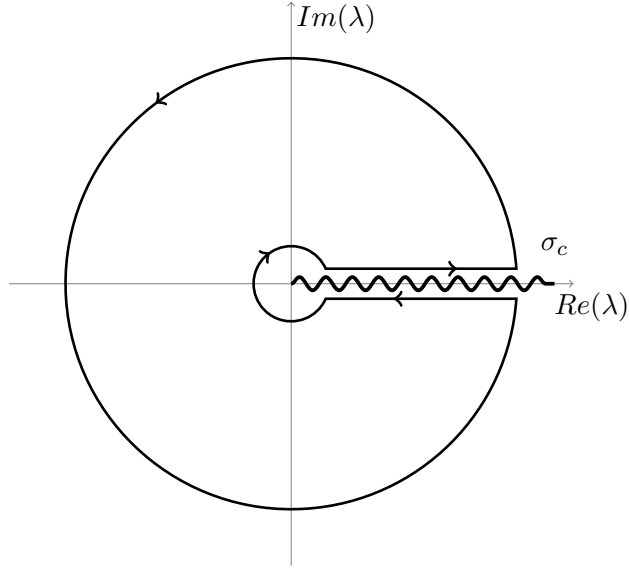
\begin{figure}[h]
	\centering
	\begin{tikzpicture}
		%Hole of the keyhole
		\def\gap{0.4}
		%Great circle
		\def\bigradius{3}
		%Little circle
		\def\littleradius{0.5}
		% Axes
		\draw [help lines,->] (-1.25*\bigradius, 0) -- (1.25*\bigradius,0);
		\draw [help lines,->] (0, -1.25*\bigradius) -- (0, 1.25*\bigradius);
		% Red path
		\draw[line width=1pt,   decoration={ markings,
			mark=at position 0.2455 with {\arrow[line width=1.2pt]{>}},
			mark=at position 0.765 with {\arrow[line width=1.2pt]{>}},
			mark=at position 0.87 with {\arrow[line width=1.2pt]{>}},
			mark=at position 0.97 with {\arrow[line width=1.2pt]{>}}},
		postaction={decorate}]
		let
		\n1 = {asin(\gap/2/\bigradius)},
		\n2 = {asin(\gap/2/\littleradius)}
		in (\n1:\bigradius) arc (\n1:360-\n1:\bigradius)
		-- (-\n2:\littleradius) arc (-\n2:-360+\n2:\littleradius)
		-- cycle;
		%Branch-cut
		\draw [line width=.5mm,style={decorate, decoration=snake}] (0,0) -- (3.5,0);
		%\draw [line width=0.25mm, red ] (0,-1) -- (2,-1) node
		% Labels
		\node at (4,-0.3){$Re(\lambda)$};
		\node at (0.6,3.53) {$Im(\lambda)$};
		\node at (3.5,.5) {$\sigma_c$};
	\end{tikzpicture}
	\caption{The keyhole contour $C_\infty$ and the branch cut corresponding to $\sigma_c$.}\label{contour_plot}
\end{figure}

Standard results on Sturm-Lioville problems \cite{Zettl:2005} yield
\begin{equation}\label{eq:Green-func}
	\mathcal{R}(z,z^\prime;\lambda) = N_\lambda \left[ \Theta(z-z^\prime)u(z;\lambda)v(z^\prime;\lambda)+\Theta(z^\prime-z)u(z^\prime;\lambda)v(z;\lambda) \right],
\end{equation}
where 
$u(z;\lambda)$ is a solution of $Lu=\lambda u$, $\lambda\in\mathbb{C}$, such that there exists $z_0>0$ for which $u\in L^2((0,z_0);dz)$, while $v(z;\lambda)$ is a solution of $Lv=\lambda v$ such that there exists $z_1>0$ for which $L^2((z_1,\infty);dz)$. The normalization in Equation \eqref{eq:Green-func} is 
\begin{equation}\label{Eq: normalization}
	N_\lambda = -\dfrac{1}{W_z[u(\cdot,\lambda),v(\cdot,\lambda)]},
\end{equation}
where $W_z$ is the Wronskian between $u(z,\lambda)$ and $v(z,\lambda)$. Starting from $u(z,\lambda)$, the analysis of Section \ref{Sec: Data} and Equation \eqref{eq:solution_massive} in particular entail that its r\^{o}le is played by
\begin{equation}
	u(z,\lambda) = \epsilon  \widehat{\Phi}_1(\lambda,z) + \zeta \widehat{\Phi}_2(\lambda,z),
\end{equation}
where the r\^{o}le of $\lambda$ is played by $q^2$ while $\epsilon = \widehat{\rho}$ and $\zeta = \dfrac{\widehat{\rho}}{c[q^2-m_b^2]}$. Focusing on $v(z,\lambda)$, its r\^{o}le is played by
\begin{equation}
	\left\{\begin{array}{ll}
		\widehat{\Phi}^\uparrow(z,\lambda) = \sqrt{z} H_\nu^{(1)}(\sqrt{\lambda}z) & \textrm{if}\;Im(\lambda)>0\\
		\widehat{\Phi}^\downarrow(z,\lambda) = \sqrt{z} H_\nu^{(2)}(\sqrt{\lambda}z) & \textrm{if}\;Im(\lambda)<0
	\end{array}\right.,
\end{equation}
where $H_\mu^{(i)}(\sqrt{\lambda}z)$, $i=1,2$, is the Hankel function of first or of second kind. In order to evaluate Equation \eqref{Eq: normalization} and assuming $\nu \neq 0$ ({\it i.e.} $m^2 > -1/4$), using \cite[Eq. 10.4.7 \& 10.4.8]{NIST} it holds
 \begin{equation}
	\widehat{\Phi}^\uparrow(\lambda,z) = \alpha  \widehat{\Phi}_1(\lambda,z) + \beta  \widehat{\Phi}_2(\lambda,z), 
\end{equation}
and
\begin{equation}
	\widehat{\Phi}^\downarrow (\lambda, z) = \bar{\alpha} \widehat{\Phi}^1 (\lambda,z) + \bar{\beta}\widehat{\Phi}^2(\lambda,z),
\end{equation}
where
\begin{align}
	&\alpha = i\sqrt{\dfrac{2}{\pi \sin(\pi \nu)}}   q^\nu e^{i \pi \nu}, \quad
	&\beta =  i\sqrt{\dfrac{2}{\pi \sin(\pi \nu)}} q^{-\nu}.
\end{align}
Since $W_z[\widehat{\Phi}_1(\lambda,z),\widehat{\Phi}_2(\lambda,z)]=\sin(\pi\nu)$, it descends
\begin{subequations}\label{eq:Wronskians}
	\begin{align}
		W_z[u(z,\lambda),\widehat{\Phi}_{\underline{k}}^\uparrow(z)]= (\epsilon\beta-\zeta\alpha)\sin(\pi\nu),\\
		W_z[u(z,\lambda),\widehat{\Phi}_{\underline{k}}^\downarrow(z)]= (\epsilon\bar{\beta}-\zeta\bar{\alpha})\sin(\pi\nu).
	\end{align}
\end{subequations}
Inserting these results into Equation \eqref{eq:Green-func}, we can write the following integral representation of the Dirac delta distribution, see \cite{Stakgold} or \cite[App. C1]{DF16}
\begin{equation}\label{eq:resolution_of_identity_integral}
	-\dfrac{1}{2\pi i}\oint_{C_\infty} d\lambda\,\mathcal{R}(z,z^\prime;\lambda) = \delta(z-z^\prime),
	% \int_{\sigma_c} d \lambda  \psi(\lambda,z)\bar{\psi}(\lambda,z^\prime) = 
\end{equation}
where $C_\infty$ is an infinitely large keyhole contour in the $\lambda$ complex plane with a counterclockwise orientation, see Figure \ref{contour_plot}. We stress the existence of a branch cut on the positive part of the real axis which reflects that all $\lambda\in\mathbb{R}^+$ lie in the continuous spectrum of the operator $L$, see Equation \eqref{Eq: ODE_G}.

In order to evaluate Equation \eqref{eq:resolution_of_identity_integral}, we apply Jordan's lemma to conclude that the contribution coming from the larger circle vanishes. As a consequence 
\begin{multline}\label{eq:complex_integral}
	-\dfrac{1}{2\pi i}\int_0^\infty d\lambda \mathcal{R}(z,z^\prime;\lambda) = \\
	-\dfrac{1}{2\pi i}\int_0^\infty d | \lambda | \ \lim_{\varepsilon \rightarrow 0^+} \left[ \mathcal{R}(z,z^\prime;\lambda+i\varepsilon)-\mathcal{R}(z,z^\prime;\lambda-i\varepsilon) \right] = - \dfrac{1}{2\pi i}\int_0^\infty d | \lambda | \Delta\mathcal{R},
\end{multline}
where
%\begin{equation}
%	\Delta \Lambda = -\sqrt{z z'}\pi i \frac{[A \widehat \Phi_1(z q) + B \hat \Phi_2(z q)][A \hat \Phi_1(z' q) + B \hat \Phi_2(z' q)]}{A^2 q^{-2 \nu} + B^2 q^{2 \nu} + 2AB \cos(\pi \nu) } 
%\end{equation} 
%with $ A = \hat \rho$ and $B =  - \hat{F}$, $c (q^2 - m_{\rm b}^2) B = A$. Substituting the explicit expressions of $A$ and $B$ yields
\begin{equation}\label{eq:delta_green}
	\Delta \mathcal{R} = -i \sqrt{\frac{2sin(\pi\nu)}{\pi}} \frac{[c (\lambda - m_{\rm b}^2) \widehat \Phi_1(z,\lambda) + \widehat \Phi_2(z,\lambda)][c (\lambda - m_{\rm b}^2) \widehat \Phi_1(z^\prime,\lambda) +  \widehat \Phi_2(z',\lambda)]}{c^2 (\lambda - m_{\rm b}^2)^2 \lambda^{-\nu} + \lambda^\nu + 2c (\lambda - m_{\rm b}^2) \cos(\pi \nu) }.
\end{equation} 
A direct inspection of Equation \eqref{eq:resolution_of_identity_integral} yields
\begin{multline}\label{eq-resolution-id}
	\delta(z, z')  = - \dfrac{1}{2\pi i} \int_0^\infty \! d q \, 2 q \ \Delta \mathcal{R}=\\
	= \sqrt{\frac{2sin(\pi\nu)}{\pi}} \int_0^\infty \! d q \, q  \frac{[c (q^2 - m_{\rm b}^2) \widehat \Phi_1(z q) + \widehat \Phi_2(z q)][c (q^2 - m_{\rm b}^2) \widehat \Phi_1(z' q) +  \widehat \Phi_2(z' q)]}{c^2 (q^2 - m_{\rm b}^2)^2 q^{-2 \nu} + q^{2 \nu} + 2c (q^2 - m_{\rm b}^2) \cos(\pi \nu) },
\end{multline}
where we reinstated $q$ via the defining relation $\lambda=q^2$. To conclude, using this last expression in Equation \eqref{eq:delta-rep}, we end up with
\begin{multline}\label{eq-G-hat}
		\widehat{G}_{k}^{\mathbb{H}}(z,z^\prime) = \frac{\sqrt{ \pi} \Gamma ( \frac{d}{2} )}{(2 \pi)^{\frac{d}{2}} \Gamma \left( \frac{d-1}{2} \right)} \sqrt{\frac{2sin(\pi\nu)}{\pi}}  \times \\ 
	\times \frac{[c (q^2 - m_{\rm b}^2) \widehat \Phi_1(z,q) + \widehat \Phi_2(z,q)][c (q^2 - m_{\rm b}^2) \widehat \Phi_1(z',q) +  \widehat \Phi_2(z',q)]}{c^2 (q^2 - m_{\rm b}^2)^2 q^{-2 \nu} + q^{2 \nu} + 2c (q^2 - m_{\rm b}^2) \cos(\pi \nu) } .
\end{multline}

Putting all the elements together, we have that the causal propagator with Wentzell boundary conditions takes the form
\begin{align}
	G^{\mathbb{H}}(x,x^\prime) 	& = \lim\limits_{\varepsilon \rightarrow 0^+} \sqrt{\frac{2}{\pi}} \int_0^\infty dk k \left(\dfrac{k}{r} \right)^{\frac{d-3}{2}} J_{\frac{d-3}{2}}(kr)\int_0^\infty dq q \frac{\sin(\sqrt{q^2+k^2}(t-t^\prime-i\varepsilon))}{\sqrt{q^2+k^2}} \widehat{G}^{\mathbb{H}}_{k}(z,z^\prime)
 \nonumber \\
	& = \lim\limits_{\varepsilon \rightarrow 0^+} \sqrt{\frac{2}{\pi}} \int_0^\infty dk k \left(\dfrac{k}{r} \right)^{\frac{d-3}{2}} J_{\frac{d-3}{2}}(kr)\int_k^\infty d\omega \sin(\omega(t-t^\prime-i\varepsilon)) \widehat{G}^{\mathbb{H}}_{k}(z,z^\prime).
\end{align}

\noindent We remark that the initial conditions in Equation \eqref{eq:commutator_function_sturm_liouville} are automatically implemented.

\paragraph{Step 3: Construction of the bulk ground state $\lambda_2$ -- } We are now in a position to write down directly the ground state in Minkowski half-space in terms of \eqref{eq-G-hat} as
\begin{align}
\lambda_2^{\mathbb{H}}(x,x') = \lim\limits_{\varepsilon \rightarrow 0^+} \sqrt{\frac{1}{2 \pi}} \int_0^\infty dk k \left(\dfrac{k}{r} \right)^{\frac{d-3}{2}} J_{\frac{d-3}{2}}(kr)\int_0^\infty dq q \frac{e^{-i\sqrt{q^2+k^2}(t-t^\prime-i\varepsilon)}}{\sqrt{q^2+k^2}} \widehat{G}^{\mathbb{H}}_{k}(z,z^\prime),
\end{align}
where we have selected only positive frequencies. Hence, in PAdS$_{d+1}$ the ground state Wightman function in the bulk is
\begin{align}
\lambda_2(x,x') = \lim\limits_{\varepsilon \rightarrow 0^+} (z z')^{\frac{d-1}{2}}  \sqrt{\frac{1}{2 \pi}} \int_0^\infty dk k \left(\dfrac{k}{r} \right)^{\frac{d-3}{2}} J_{\frac{d-3}{2}}(kr)\int_0^\infty dq q \frac{e^{-i\sqrt{q^2+k^2}(t-t^\prime-i\varepsilon)}}{\sqrt{q^2+k^2}} \widehat{G}^{\mathbb{H}}_{k}(z,z^\prime).
\label{TwoPtFn}
\end{align}

Having constructed the two-point correlation function for the ground state in Equation \eqref{TwoPtFn} combined with Equation \eqref{eq-G-hat}, we are now in position to infer a notable consequence. As a matter of fact, in view of the general analysis in \cite{Sahlmann:2000fh} and of the results on the propagation of singularities on asymptotically AdS spacetimes proven in \cite{Dappiaggi:2020yxg}, we can conclude that $\lambda_2$ is a bi-distribution of Hadamard form in PAdS$_{d+1}$, both locally and globally, in the sense of \cite{Dappiaggi:2017wvj}.

\section{Bulk-boundary and boundary-boundary propagators}

As next step we observe that, without entering into the technical details, far from the scope of this work, the singular structure of the bi-distribution $\lambda_2^{\mathbb{H}}$ is such that it is always possible to extend it to $\partial\mathbb{H}^{d+1}$. As a consequence, we can restrict one or both of its entries to lie on $\partial\mathbb{H}^{d+1}$, hence giving rise to what is known in the literature as the {\em bulk-to-boundary} $\lambda_{2B\partial'}^{\mathbb{H}}$ or {\em boundary-to-boundary} $\lambda_{2\partial\partial^\prime}^{\mathbb{H}}$ two-point function. While giving an explicit expression of these propagators is elusive in the general scenario, it is instructive to focus on the case $m=0$, since concrete formulae can be written. More precisely, in this scenario all expressions derived in previous section, can be straightforwardly continued to $\partial\mathbb{H}^{d+1}$. To make this statement more precise and thus following the results of Section \ref{Sec:GroundState}, we set 
\begin{align}\label{eq:eigenfuction_massless}
\psi_q(z) & = c (q^2- m_{\rm b}^2) \widehat \Phi_1(z,\sqrt{q^2}) + \widehat \Phi_2(z,\sqrt{q^2}), 
\end{align}
and
\begin{align}
\mathcal{N}_{q^2} \dot = \left(  \frac{ \Gamma ( \frac{d}{2} ) \left(2\sin(\pi\nu)\right)^{1/2}}{(2 \pi)^{\frac{d}{2}} \Gamma \left( \frac{d-1}{2} \right) \left(c^2 (q^2 - m_{\rm b}^2)^2 q^{-2 \nu} + q^{2 \nu} + 2c (q^2 - m_{\rm b}^2) \cos(\pi \nu) \right)} \right)^{1/2}.
\end{align}

We start from the bulk-to-bulk two-point function as per Equation \eqref{eq-G-hat}, here reported for convenience:
\begin{align}
{\lambda}_2^{\mathbb{H}}(x, x') = \lim\limits_{\varepsilon \rightarrow 0^+} \sqrt{\frac{1}{2 \pi}} \int_0^\infty dk k \left(\dfrac{k}{r} \right)^{\frac{d-3}{2}} J_{\frac{d-3}{2}}(kr)\int_0^\infty dq q \frac{e^{-i\sqrt{q^2+k^2}(t-t^\prime-i\varepsilon)}}{\sqrt{q^2+k^2}} \widehat{\mathcal{G}}^{\mathbb{H}}_{k}(z,z^\prime).
\end{align}

\noindent A direct inspection of Equation \eqref{eq:eigenfuction_massless} shows that, being $m=0$, it holds that
$$\psi_q(0) = \widehat \Phi_2(0,\sqrt{q^2}),$$
which, in combination with Equation \eqref{eq-G-hat}, entails that we can construct the boundary-to-boundary two-point correlation function simply setting $z=z^\prime=0$. Denoting it henceforth by $\lambda_{2 \partial \partial'}^{\mathbb{H}}$, it reads
\begin{align}
{\lambda}_{2\partial\partial^\prime}^{\mathbb{H}} = \lim\limits_{\varepsilon \rightarrow 0^+} \sqrt{\frac{1}{2 \pi}} \int_0^\infty dk k \left(\dfrac{k}{r} \right)^{\frac{d-3}{2}} J_{\frac{d-3}{2}}(kr)\int_0^\infty dq q \frac{e^{-i\sqrt{q^2+k^2}(t-t^\prime-i\varepsilon)}}{\sqrt{q^2+k^2}} |\mathcal{N}_{q^2} \widehat \Phi_2(0,q)|^2.
\end{align}
Similarly one can wonder what happens if only one of the two legs of $\lambda_2^{\mathbb{H}}$ is restricted to the boundary. This gives rise to the bulk-to-boundary two-point correlation function, which we denote by $\lambda_{2 {\rm B} \partial'}^{\mathbb{H}}$. Setting $z^\prime=0$ it reads
\begin{align}
{\lambda}_{2 {\rm B} \partial'}^{\mathbb{H}}= \lim\limits_{\varepsilon \rightarrow 0^+} \sqrt{\frac{1}{2 \pi}} \int_0^\infty dk k \left(\dfrac{k}{r} \right)^{\frac{d-3}{2}} J_{\frac{d-3}{2}}(kr)\int_0^\infty dq q \frac{e^{-i\sqrt{q^2+k^2}(t-t^\prime-i\varepsilon)}}{\sqrt{q^2+k^2}} |\mathcal{N}_{q^2}|^2  \psi_q(z) \overline{ \widehat \Phi_2(0,q)}.
\end{align}

To conclude the section, we remark that a natural question is whether the singular structure of $\lambda_{2 \partial \partial'}^{\mathbb{H}}$ is related to that of $\lambda_2^{\mathbb{H}}$. Since answering this question would require a detailed and lengthy mathematical analysis, we content ourselves to commenting on this issue. First of all we recall that there exists a global and a local version of the so-called Hadamard condition for a bi-distribution, say $\omega_2$, see \cite{Khavkine:2014mta}. These are equivalent provided that $\omega_2$ is a weak bi-solution of a linear, second order, hyperbolic partial differential equation, such that its antisymmetric part is proportional to the associated retarded-minus-advanced fundamental solution. In the case in hand, one can employ techniques proper of microlocal analysis to prove that the global singular structure of $\lambda_{2 \partial \partial'}^{\mathbb{H}}$ is consistent with the Hadamard condition on $\mathbb{R}^d\simeq\partial\mathbb{H}^{d+1}$. Yet it is not of local Hadamard form since, even though, this bi-distribution is per construction a bi-solution of the Klein-Gordon equation on the boundary, its anti-symmetric part does not proportional to the retarded-minus-advanced fundamental solution of $\Box_\eta^{(d)}-m^2_b$, as one can infer from direct inspection.

%\appendix

\section{Support properties of the Green functions}
\label{sec:Support}

The construction outlined in the main body of this work guarantees that $G^{\mathbb{H}}$ is a bi-solution of the equation of motion ruled by $P_\eta$, see Equation \eqref{eq:KG_boundary_problem_singular}, supplemented by boundary conditions of Wentzell type. Yet, there is no a priori guarantee that $G^{\mathbb{H}}$ codifies the standard causal properties of the advanced and of the retarded fundamental solutions of a wave-like operator on a globally hyperbolic spacetime, namely that, for every $f\in C^\infty_0(\mathring{\mathbb{H}}^{d+1})$, $\textrm{supp}(G^{\mathbb{H}}(f))\subseteq J^+(\textrm{supp}(f))\cup J^-(\textrm{supp}(f))$, where $J^\pm$ indicate the causal future ($+$) and past ($-$) of a subset of $\mathbb{H}^{d+1}$ endowed with the Minkowski metric. In the following we use a so-called energy estimate to prove that such structural property holds true in the case in hand for a bulk massless scalar field, see Equation \eqref{eq:KG_boundary_problem_singular} with $m=0$. Observe that, since PAdS$_{d+1}$ is conformally related to $(\mathbb{H}^{d+1},\eta)$ and thus since they share the same causal structure, the support properties of $G^{\mathbb{H}}$ are inherited automatically by the causal propagator $G$ associated to the operator $P$, see Equation \eqref{KG-PAdS}, via the defining relation $G=(zz^\prime)^{\frac{d-1}{2}}G^{\mathbb{H}}$. In order to use energy estimates, it is convenient to reformulate the problem in hand in terms of an extended Hilbert space. More precisely, let us consider

\begin{equation}
	\mathcal{H} = L^2(\mathring{\mathbb{H}}^d) \oplus L^2(\mathbb{R}^{d-1}),
\end{equation}
with inner product
\begin{equation}
	(u,v)_\mathcal{H} := (u_1,v_1)_{L^2(\mathring{\mathbb{H}}^d)} + (u_2,v_2)_{L^2(\mathbb{R}^{d-1})}. 
\end{equation}
for $u = \begin{pmatrix} u_1 \\ u_2 \end{pmatrix}, v = \begin{pmatrix} v_1 \\ v_2 \end{pmatrix} \in \mathcal{H} $. Here $\mathring{\mathbb{H}}^d$ and $\mathbb{R}^{d-1}$ have to be interpreted respectively as slices of $\mathring{\mathbb{H}}^{d+1}$ and of $\partial\mathbb{H}^{d+1}$ at constant time $t$.

Let $\mathcal{G} := G^{\mathbb{H}} \oplus G^{\partial \mathbb{H}}$ be the direct sum of the bulk-to-bulk and the boundary-to-boundary advanced-minus-retarded propagators for the massless Klein-Gordon operator with Wentzell boundary conditions, the latter being the pullback of the former to the boundary. Let $u(t) = [\mathcal{G}(h)](t) = (\Phi(t),F(t))$, with $h \in \mathcal{C}^\infty_0(\mathring{\mathbb{H}}^{d+1}) \times\mathcal{C}^\infty_0(\mathbb{R}^d)$ be a solution of the problem in hand, where we highlight the dependence from the time coordinate $t$. Furthermore, let $K \subset\{0\} \times \mathring{\mathbb{H}}^d$ be a compact set in $\mathbb{H}^{d+1}$ and we assume $f$ to be chosen so to have vanishing initial data:
$$u(0)|_{K} = 0, \quad \dot{u}(0)|_{K} = 0, \quad F(0)|_{K \cap \partial\mathbb{H}^{d+1}}=0 \quad \textit{and} \quad \dot{F}(0)|_{K \cap \partial\mathbb{H}^{d+1}} = 0.$$

Consider a time slice $M_t = \{t\} \times \mathbb{H}^d$, $t \in \mathbb{R}$, and let $K_t = M_t \cap J(K)$, where $J(K) := J^+(K) \cup J^-(K)$. To simplify the notation, we call $\partial_b K_t\doteq\partial\mathbb{H}^{d+1} \cap \partial K_t$,  $\partial^\circ K_t\doteq\mathring{\mathbb{H}}^{d+1} \cap \partial K_t$. We consider the energy functional $E[u](t) = E^\circ[\Phi](t)+E_\partial [F](t)$ with
\begin{equation}\label{eq:PAdS_bulk_E}
	E^\circ[\Phi](t) = \frac{1}{2} \left[ \| \Phi \|^2_{L^2(K_t)} +  \| \dot{\Phi} \|^2_{L^2(K_t)} +  \| \nabla \Phi \|^2_{L^2(K_t)} \right]
\end{equation}
\begin{equation}\label{eq:PAdS_bulk_E_2}
	E_\partial [F](t) = \frac{1}{2} \left[ \| F \|^2_{L^2(\partial_b K_t)} +  \| \dot{F} \|^2_{L^2(\partial_b K_t)} +  \| \nabla_\partial F \|^2_{L^2(\partial_b K_t)} \right],
\end{equation}
where $\nabla_\partial$ indicates the covariant derivative pulled back to $\partial_b K_t$.
Observe that, for any solution $u$ of the problem in hand, one can show that both Equation \eqref{eq:PAdS_bulk_E} and  \eqref{eq:PAdS_bulk_E_2} are finite \cite{DDF18}. To fix notation, in the following, by $(,)$ we shall denote the inner product of $L^2(K_t)$. 
The time variation of Equation \eqref{eq:PAdS_bulk_E} reads
\begin{equation}\label{eq:pads_bulk_derivative}
	\begin{split}
		\frac{d}{dt} E^\circ[\Phi](t) = & (\Phi,\dot{\Phi}) + (\dot \Phi,\ddot{\Phi}) + (\nabla \Phi, \nabla \dot{\Phi}) \\
		& - \frac{1}{2} \left[ \| \gamma \Phi \|^2_{L^2(\partial^\circ K_t)} +  \| \gamma \dot{\Phi} \|^2_{L^2(\partial^\circ K_t)} +  \| \gamma \nabla_\partial \Phi \|^2_{L^2(\partial^\circ K_t)}\right],\\
		& - \frac{1}{2} \left[ \| F \|^2_{L^2(\partial_b K_t)} +  \| \dot{F} \|^2_{L^2(\partial_b K_t)} +  \| \gamma \nabla_\partial \Phi \|^2_{L^2(\partial_b K_t)} \right],
	\end{split}
\end{equation}
where, with a slight abuse of notation, we indicate with $\gamma$ both the restriction map to $\partial^\circ K_t$ as well as to $\partial_b K_t$. In both cases one can show that $\gamma$ is well-defined \cite{DDF18} and \cite{Dappiaggi:2020yxg}. The time derivative of $E_\partial[\Phi](t)$ in Equation \eqref{eq:PAdS_bulk_E_2} is 
\begin{equation}\label{eq:pads_boundary_derivative}
	\frac{d}{dt} E_\partial[\Phi](t) = (F, \dot{F})_{L^2(\partial_b K_t)} + (\dot{F},\ddot{F})_{L^2(\partial^\circ K_t)} +  (\nabla_\partial F, \nabla_\partial \dot{F})_{L^2(\partial_b K_t)}.
\end{equation}
Focusing first on Equation \eqref{eq:pads_bulk_derivative} and integrating by parts, we obtain
\begin{equation}
	(\nabla \Phi, \nabla \dot{\Phi}) = - (\bigtriangleup \Phi, \dot{\Phi}) + \int_{\partial_b K_t} \gamma \dot \Phi \overline{\gamma \nabla_\perp \Phi} d\mu_{\partial_b K_t} + \int_{\partial_b K_t} \gamma \dot \Phi \overline{\gamma \nabla_\perp \Phi} d\mu_{\partial^\circ K_t}.
\end{equation}
where $\nabla_\perp$ is the covariant derivative along the direction normal to the boundary. Using Equation \eqref{KG-PAdS} and the inequality $(v,w)\leq\frac{1}{2}(\|v\|^2+\|w\|^2)$ valid for any $v,w$ lying in a real vector space endowed with a scalar product $(,)$, yields the following chain of inequalities
\begin{equation}\label{eq:bound_pads_bulk}
	\begin{split}
		\frac{d}{dt} E^\circ[\Phi](t)  = (\Phi,\dot{\Phi})+ (\gamma \dot{\Phi},\gamma \nabla_\perp \Phi)_{L^2(\partial^\circ K_t)} + (\gamma \dot{\Phi},\rho)_{L^2(\partial_b K_t)} + \\
		- \frac{1}{2} \left[ \| \gamma \Phi \|^2_{L^2(\partial^\circ K_t)} +  \| \gamma \dot{\Phi} \|^2_{L^2(\partial^\circ K_t)} +  \| \gamma \nabla_\partial \Phi \|^2_{L^2(\partial^\circ K_t)}\right]\\
		- \frac{1}{2} \left[ \| F \|^2_{L^2(\partial_b K_t)} +  \| \dot{F} \|^2_{L^2(\partial_b K_t)} +  \| \gamma \nabla_\partial \Phi \|^2_{L^2(\partial_b K_t)} \right] \\
		\leq \frac{\| \Phi \|^2_{L^2(K_t)}}{2} + \frac{\| \dot{\Phi}\|^2_{L^2(K_t)}}{2} + \frac{1}{2} \left[ \| \gamma \dot{\Phi} \|^2_{L^2(\partial^\circ K_t)} + \| \gamma \nabla_\perp \Phi \|^2_{L^2(\partial^\circ K_t)} \right] + \\
		+ \frac{1}{2} \left[ \| \dot{F} \|^2_{L^2(\partial_b K_t)} + \| 	\rho \|^2_{L^2(\partial_b K_t)} \right] + \\
		- \frac{1}{2} \left[ \| \gamma \Phi \|^2_{L^2(\partial^\circ K_t)} +  \| \gamma \dot{\Phi} \|^2_{L^2(\partial^\circ K_t)} +  \| \gamma \nabla_\partial \Phi \|^2_{L^2(\partial^\circ K_t)}\right]\\
		- \frac{1}{2} \left[ \| F \|^2_{L^2(\partial_b K_t)} +  \| \dot{F} \|^2_{L^2(\partial_b K_t)} + \| \rho \|^2_{L^2(\partial_b K_t)} \right] \leq \\
		\leq \frac{1}{2} \| \Phi \|^2_{L^2(K_t)} +\frac{1}{2} \| \dot{\Phi} \|^2_{L^2(K_t)} \leq E^\circ[u](t),
	\end{split}
\end{equation}
where we have implicitly taken into account the equation of motion and the possibility to choose $K_t$ in such a way that the restriction of $\Phi$ and of its derivatives on $\partial^\circ K_t$ vanishes identically. Focusing now on Equation \eqref{eq:PAdS_bulk_E_2} and proceeding as for the previous term, we can bound the time variation of $E_\partial[F](t)$ as follows
\begin{equation}
	\begin{split}
		\frac{d}{dt}E_\partial[F](t) \leq \| F \|^2_{L^2(\partial_b K_t)} + \| \dot{F} \|^2_{L^2(\partial_b K_t)} - (m_b F, \dot{F})_{L^2(\partial_b K_t)} + \frac{1}{2c} \| \rho \|^2_{L^2(\partial_b K_t)}+ &\\
		+ \frac{1}{2c} \| \dot{F} \|^2_{L^2(\partial_b K_t)}-\frac{1}{2} \left[ \| F \|^2_{L^2(\partial_b K_t)} + \| \dot{F} \|^2_{L^2(\partial_b K_t)} + \| \nabla_\partial F \|^2_{L^2(\partial_b K_t)} \right] \leq \\
		\leq C \left[ \| F \|^2_{L^2(\partial_b K_t)} + \| \dot{F} \|^2_{L^2(\partial_b K_t)}\right] + \frac{1}{2c}\| \rho \|^2_{L^2(\partial_b K_t)},
	\end{split}
\end{equation}
with $C = 1 + \frac{1}{2|c|}+\frac{m_b}{2}$.
Since $c<0$ per hypothesis, then we can estimate the last expression from above as
\begin{equation}\label{eq:bound_pads_boundary}
	\begin{split}
		E_\partial[F](t) \leq  C \left[ \| F \|^2_{L^2(\partial_b K_t)} + \| \dot{F} \|^2_{L^2(\partial_b K_t)}\right] \leq 2 C E_\partial[F](t),
	\end{split}
\end{equation}
Combining Equations \eqref{eq:bound_pads_boundary} and Equation \eqref{eq:bound_pads_bulk} yields
\begin{equation}\label{Eq: estimate}
\dfrac{d}{dt}E[u](t) \leq 2C E[u](t).
\end{equation}
Therefore, applying Gronwall's lemma, we obtain that $E[u](t) \leq e^{2C t} E[u](0)$. Being $E[u](t)$ positive, the vanishing of $u$ and, therefore, of $E[u]$ on $K$ yields the sought result. Indeed, if $E[u](0)=0$ Equation \eqref{Eq: estimate} entails that $E[u](t)=0$ in $K_t$. Hence we conclude that the solution $u(t)=(\Phi(t),F(t))$ vanishes on $K_t$ and therefore that $G$ has the desired support property.

\section{Final remarks}
\label{sec:Rem}

In this paper we have studied a class of real Klein-Gordon field theories in the Poincar\'e fundamental domain of anti de-Sitter spacetime in $d+1$ dimensions with dynamical Wentzell boundary conditions, which admit the definition of a ground state. This implies that the boundary conditions at conformal infinity depend explicitly on second order time derivatives of the field, and indeed take the form of non-homogeneous wave equations for a boundary field, where the source term is given by the trace $\gamma_1 \Phi = (\partial_z \Phi)|_{z = 0}$ of the bulk field.

The main task has been to obtain explicit expressions for the ground state Wightman two-point functions as mode expansions, which we achieve in Equation \eqref{TwoPtFn} by exploiting the symmetry of spacetime and reducing the problem to finding the Green operator of an associated Sturm-Liouville problem, see Equation \eqref{eq-G-hat}. As a by-product of \cite{Dappiaggi:2021wtr, Dappiaggi:2020yxg} we can also infer that the Wightman two-point function enjoys the Hadamard property. 

As expected, the anti-symmetric part of the two-point function is proportional to what is expected to be the causal propagator of the theory, and we verify that this anti-symmetric part satisfies the desired boundary conditions at $t = 0$ and Wentzell boundary conditions at the conformal boundary. Furthermore, in the massless ($m^2 = 0$) Klein-Gordon case we can show that the support properties of the advanced-minus-retarded propagator are the standard ones. However, proving an analogous result for the massive case remains elusive.

In the massless case it is also possible to naturally define by restriction a boundary state for the boundary quantum field theory, which appears naturally as the boundary-to-boundary propagator of a matrix-valued two-point function with further bulk-to-bulk and bulk-to-boundary propagators.

As a final comment, we should mention that with the ground state expression \eqref{TwoPtFn} that we have obtained it is an easy task to obtain a KMS two-point function at positive temperature. Indeed, it is easy to see that if \eqref{TwoPtFn} defines a ground state for the theory, then
\begin{align}
\lambda_2^\beta(x,x') & = \lim\limits_{\varepsilon \rightarrow 0^+} (z z')^{\frac{d-1}{2}}  \sqrt{\frac{1}{2 \pi}} \int_0^\infty dk k \left(\dfrac{k}{r} \right)^{\frac{d-3}{2}} J_{\frac{d-3}{2}}(kr) \nonumber \\
& \times \int_0^\infty dq q \frac{\left( 1 - e^{-\beta \sqrt{q^2+k^2}} \right)^{-1}}{\sqrt{q^2+k^2}} \widehat{G}^{\mathbb{H}}_{k}(z,z^\prime) \left( e^{-i\sqrt{q^2+k^2}(t-t^\prime-i\varepsilon)} + e^{-\beta \sqrt{q^2+k^2}} e^{i\sqrt{q^2+k^2}(t-t^\prime-i\varepsilon)} \right)
\label{KMS}
\end{align}
defines correspondingly a KMS state in equilibrium at temperature $ T = 1/\beta > 0$.

\section*{Acknowledgments}

CD is grateful to the INFN Sezione di Pavia for travel support during the realization of this work. BAJ-A thanks the hospitality of the University of Pavia and INFN Sezione di Pavia, where part of this research was undertaken. BAJ-A is supported by a CONACYT Postdoctoral Research Fellowship, and acknowledges in addition the support of CONACYT project 140630 and UNAM-DGAPA-PAPIIT grant IG100120.

%\bibliographystyle{unsrt}
%\bibliography{biblio}

\end{document}